\documentclass[prb,twocolumn]{revtex4}

\usepackage{graphicx}
\usepackage{dcolumn}
\usepackage{bm}

\begin{document}

\title{Monte Carlo simulation of equilibrium L1$_0$ ordering in FePt nanoparticles}

\author{R.V. Chepulskii$^{1,2,*}$, J. Velev$^{1}$ and W.H. Butler$^{1}$}

\affiliation{$^{1}$Center for Materials for Information
Technology, University of Alabama, Box 870209, Tuscaloosa, Alabama 35487-0209, USA\\
$^{2}$Department of Solid State Theory, Institute for Metal
Physics, N.A.S.U., Vernadsky 36, Kyiv-142, UA-03680, Ukraine}

\date{Received 7 September 2004; accepted 10 October 2004 by J. Appl. Phys.}

\begin{abstract}

First, second and third nearest neighbor mixing potentials for
FePt alloys, were calculated from first principles using a
Connolly-Williams approach. Using the mixing potentials obtained
in this manner, the dependency of equilibrium L1$_0$ ordering on
temperature was studied for bulk and for a spherical nanoparticle
with 3.5nm diameter at equiatomic composition by use of Monte
Carlo simulation and the analytical ring approximation. The
calculated order-disorder temperature for bulk (1495-1514 K) was
in relatively good agreement (4\% error) with the experimental
value (1572K). For nanoparticles of finite size, the (long range)
order parameter changed continuously from unity to zero with
increasing temperature. Rather than a discontinuity indicative of
a phase transition we obtained an inflection point in the order as
a function of temperature. This inflection point occurred at a
temperature below the bulk phase transition temperature and which
decreased as the particle size decreased.  Our calculations
predict that 3.5nm diameter particles in configurational
equilibrium at 600$^{\circ}$C (a typical annealing temperature for
promoting L1$_0$ ordering) have an L1$_0$ order parameter of 0.83
(compared to a maximum possible value equal to unity). According
to our investigations, the experimental absence of (relatively)
high L1$_0$ order in 3.5nm diameter nanoparticles annealed at
600$^{\circ}$C or below is primarily a problem of \emph{kinetics}
rather than equilibrium.

\end{abstract}

\pacs{1}

\maketitle

\section{Introduction}

Self-assembled, monodispersed FePt nanoparticles are being
intensively investigated for possible future application as an
ultra-high density magnetic storage medium. In order to be useful
as a storage medium, these particles, because of their extremely
small volume, $V$,  must have sufficiently high magnetic
anisotropy, $K_u$, to withstand thermal fluctuations of the
direction of magnetization. This requires values of the thermal
stability factor, $(K_u V)/(k_\texttt{B} T)$, of approximately 50.
The particles are usually produced by a ``hot soap'' process that
yields a disordered fcc solid solution alloy (e.g. Ref.
\onlinecite{Sun}). Such particles are not useful for information
storage because they are superparamagnetic at room temperature due
to their low magnetic anisotropy.

Typically, the particles are annealed at a temperature
$T\simeq600^\circ$C in order to induce an ordered L1$_0$
phase\cite{Takahashi03,Sint}.  The layered L1$_0$ phase\cite{L10}
is known from studies of bulk alloys to have an extremely high
magnetic anisotropy ($K_u\cong7\times10^7$ erg/cm$^3$).  This
value of magnetic anisotropy would provide a sufficiently large
thermal stability factor to make 3.5nm diameter particles viable
for information storage.

Unfortunately, it appears to be difficult to a achieve a high
degree of long range atomic order in FePt \emph{nanoparticles}
with $\lesssim$4nm diameter by annealing at
$T\lesssim$600$^{\circ}$C (e.g. Ref.\onlinecite{Takahashi03}). One
can consider two possible reasons for the fact that it has not
been possible to obtain well ordered small particles. First, the
observed order may be low because the particle is \emph{not} in
its equilibrium state due to the slow kinetics at low
temperatures. Alternatively, the \emph{equilibrium} order itself
may be low even at relatively low temperatures because of the
small size of nanoparticles. The latter explanation was suggested
in Ref. \onlinecite{Takahashi03}. There, the order-disorder phase
transition temperature was estimated to decrease with decrease of
particle size. For particle sizes less than 1.5 nm in diameter,
the phase transition temperature was found to be below the typical
annealing $T\simeq600^{\circ}$C. Therefore, particles of diameter
less than 1.5 nm were predicted to have no long range order in
their equilibrium state at 600$^{\circ}$C. This explanation is in
qualitative agreement with experiment. The difference between the
experimental (4nm) and theoretical (1.5nm) critical size for
disappearance of L1$_0$ order at 600$^{\circ}$C was attributed to
the neglect of nanoparticle surface effects.

From our point of view,however, the results obtained in Ref.
\onlinecite{Takahashi03} require verification because of the
limitations of the theoretical models used in that study. Namely,
the interatomic potentials in alloys usually are much more
complicated and long-ranged than the nearest neighbor
Lennard-Jones model that was used.  In addition, the
order-disorder phase transition temperature was estimated in Ref.
\onlinecite{Takahashi03} by comparing the free energies of
completely ordered and completely disordered states; whereas in
reality, the ordered state approaches (with increasing
temperature) the phase transition point being not completely
ordered. Also, the disordered state would be expected to approach
the phase transition (with decreasing temperature), not with a
completely random atomic distribution but with an atomic
distribution that has substantial short range order.   Moreover,
it is known\cite{NoFinitePhTr} that there is no formal phase
transition in a finite system.

In the present paper we utilize first
principles calculations (VASP code\cite{VASP}) together with the
Connolly-Williams\cite{CW} method and Monte Carlo (MC) simulations
(utilizing the Metropolis algorithm\cite{Metropolis}) to study the
temperature dependence of equilibrium L1$_0$ order in a spherical
FePt nanoparticle with 3.5nm diameter and equatomic composition
($c=0.5$).

\section{Results}

We consider an Fe-Pt alloy in the framework of the commonly used
two-component lattice gas model. In such a model\cite{Lee52}, two
types of atoms are distributed over the sites of a rigid crystal
lattice. The atoms are allowed to be situated only at the crystal
lattice sites and each site can be occupied by only one atom. The
atoms interact through the lattice potentials (so-called mixing
potentials) and can exchange their positions according to Gibbs
statistics.

We used the Connolly-Williams\cite{CW} method to calculate the
mixing potentials. Within this method, the energies of several
periodic atomic distributions (i.e. long-range ordered structures
called superstructures; for example L1$_0$) are calculated by
first principles methods. Then the mixing potentials are
determined by the best fit to those energies. We considered twenty
three linearly independent Fe-Pt superstructures of the same
equiatomic composition $c=0.5$. First principles calculations were
performed within the local-density approximation to
density-functional theory, using the VASP program
package.\cite{VASP} All superstructures were totally relaxed
including shape and volume relaxation of the unit cell and
individual displacements of atoms within the unit cell.
An $8\times8\times8$ mesh of $k$-points in the full Brillouin zone was employed.

The L1$_0$ superstructure was included in our first set
of first principles calculations. In this case, after atom position
relaxation, we obtained $3.848$\AA\, and $3.771$\AA\, for $a$ and $c$
lattice parameters of the corresponding tetragonal lattice,
respectively. For comparison the experimental values are
$3.847$\AA\, and $3.715$\AA.\cite{acExper} In addition, our calculated
results showed the L1$_0$ ferromagnetic superstructure to be more stable (i.e. has
lower energy) than the antiferromagnetic one in accordance with experiment.
We believe that this good correspondence between theoretical and
experimental results confirms the adequacy of our VASP first
principles calculations.

By applying the Connolly-Williams method, we
obtained 0.08769 eV, -0.03946 eV and 0.01585 eV for the first,
second and third nearest neighbor pair mixing potentials,
respectively. The average accuracy with which we fit the energy of the
twenty-three superstructures within the
Connolly-Williams method was 1.14\% per one structure.

To verify the calculated values of mixing potential, we calculated
the phase transition temperature in the \emph{bulk} FePt alloy
using these values. As a result we obtained 1495 K and 1514 K
within the analytical ring approximation\cite{Ring} and MC
simulation, respectively. The close correspondence of these values
to the experimental\cite{FePt-L10-bulk} one of 1572 K (4\% error),
demonstrates the adequacy of the calculated mixing potential.

\begin{figure}
\includegraphics{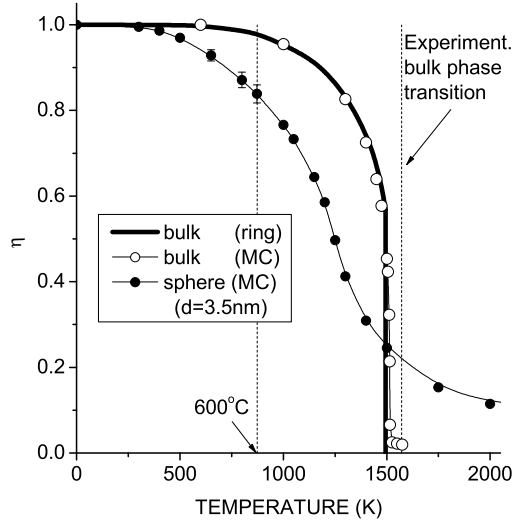}
\caption{The temperature dependence of the FePt equilibrium L1$_0$
order parameter $\eta$ in the cases of bulk ("bulk") and a
spherical nanoparticle with 3.5nm diameter ("sphere") at
equiatomic composition $c=0.5$. Two results for bulk were obtained
by Monte Carlo simulation ("MC") for parallelepiped sample
containing $N=216000$ atoms and within the analytical ring
approximation\cite{Ring} ("ring"). At simulation, the starting
configuration for each temperature was chosen to be the completely
ordered one. We applied free and periodic boundary conditions in
cases of spherical nanoparticle and parallelepiped, respectively.
For the case of a nanoparticle at 378$^{\circ}$C, 528$^{\circ}$C and
600$^{\circ}$C, the error bars correspond to dispersion of $\eta$
due to the thermodynamic fluctuations.} \label{Fig}
\end{figure}

To investigate long range order in spherical nanoparticles, we
used the calculated mixing potentials in MC simulations to
determine the temperature dependence of the equilibrium L1$_0$
order parameter $\eta$ in the case of spherical FePt nanoparticles
with 3.5nm diameter and equiatomic composition $c=0.5$. The
results are presented in Fig. \ref{Fig}.

We define the equilibrium L1$_0$ order parameter $\eta$ as the
statistical average of the maximum value among three absolute
values of "directional" order parameters $\eta_x,\eta_y,\eta_z$:
\begin{equation}\label{LRO-def}
\eta=\left\langle\max\{\left|\eta_x\right|,\left|\eta_y\right|,\left|\eta_z\right|\}\right\rangle_{\texttt{MC}},
\end{equation}
where $\eta_i\quad(i=x,y,z)$ is defined as difference between the
Fe atom concentrations at odd and even crystal planes
perpendicular to $i$-th direction,
$\left\langle\ldots\right\rangle_{\texttt{MC}}$ is the statistical
average over the MC steps. We chose such a definition of $\eta$
because of the equivalence by symmetry of the $x,y$, and $z$
directions of L1$_0$ order. In addition, one can obtain an
equivalent structure (at $c=0.5$) by changing the sign of
$\eta_i$, which results in the exchange of Fe and Pt atoms
producing a configuration that is equivalent by symmetry to the
original one. During MC simulation, we observed fluctuations that
cause transformations between these equivalent states (i.e.
fluctuations in the sign and direction of $\eta$)\cite{APB}. This
is in addition to the usual statistical fluctuations within one
such state. The L1$_0$ order parameter $\eta$, defined in Eq.
(\ref{LRO-def}) takes into account the fluctuation induced
transformations between the equivalent states\cite{SepAver}.

\section{Conclusions and discussion}

From Fig. \ref{Fig} one may conclude the following. The ring
approximation (which corresponds to bulk, i.e. an infinite
sample) clearly shows a phase transition when the order parameter
$\eta$ drops to zero. Strictly speaking, in both of the cases
considered here of finite size samples (sphere and parallelepiped)
there is no phase transition in accordance with a general
theorem\cite{NoFinitePhTr}. The order parameter $\eta$
continuously changes from unity to zero with increasing
temperature and instead of a phase transition we obtain an
inflection point in the $\eta(T)$ curve. In the case of the
parallelepiped with $216000$ atoms, the inflection point is very
similar to the phase transition.\cite{MC-PhTr}

Our calculations predict that 3.5nm diameter particles in
configurational equilibrium at 600$^{\circ}$C would have an order
parameter $\eta=0.83$ (compared to a maximum possible value of
unity). Therefore, annealing at 600$^{\circ}$C will not yield
perfect order for 3.5nm diameter particles.  Approximately 17\% of
the atoms will be on the wrong sublattices, even in equilibrium.
The dispersion of $\eta$ due to the thermodynamic fluctuations is
comparatively small (2.5\%) near annealing $T=600^{\circ}$C.

According to our investigations, the experimental absence of
(relatively) high order in nanoparticles below 600$^{\circ}$C is
primarily a \emph{kinetic} problem rather than an equilibrium one.
It should be noted that to rapidly obtain the correct equilibrium
state, we used simplified kinetics in our MC
simulation\cite{Equilibr}. Namely, we allowed \emph{any} two
randomly chosen atoms to exchange their positions \emph{without}
an additional diffusion barrier. In a real alloy, the main
mechanism of atomic diffusion is much slower because it consists
in exchange the positions between atoms and their nearest neighbor
vacancies through energy barriers. Moreover, at each temperature
we started the simulation from the completely ordered state,
whereas the actual nanoparticles are initially prepared in
disordered state and transformation from the disordered to the
ordered state may be much slower than the reverse one, especially
at low temperatures. Nevertheless, even with our simplified
kinetics, we observed a slowing down problem in approaching the
equilibrium ordered state at low temperatures. In real
nanoparticles this problem must be much worse. Kinetic
acceleration methods such as irradiation and/or addition of other
types of atoms\cite{harrell} may be useful in accelerating the
formation of long range order.

In our study we used mixing potentials obtained for
\emph{infinite} bulk alloys and used \emph{free} boundary
conditions to simulate the equilibrium configuration of finite
size particles. The presence of the surface will change the atomic
potentials in the near-surface region in comparison with bulk
potentials. Analytical estimation of such surface effects is
not straightforward and will be done elsewhere\cite{FullPaper}. In
reality, the problem of the effect of the surface on the
interatomic exchange potentials is even more complicated because
the nanoparticles of most current interest are likely to have unknown
atoms and molecules attached to their surfaces.

\section{Acknowledgments}

This research was supported by the Defense Advanced Research
Projects Agency through ONR contract N00014-02-01-0590 and by
National Science Foundation MRSEC Grant No. DMR0213985. The
authors thanks Prof. J.W. Harrell  for stimulating
discussions.


\begin{thebibliography}{*}
\bibitem{Sint}Annealing at $\sim$600$^{\circ}$C results in sintering of
nanoparticles into larger agglomerates which is not desirable for
high density magnetic recording.
\bibitem{L10} L1$_0$ is an f.c.c. tetragonal superstructure, in which
atoms of two types form layers occupying alternating (001) or
(010) or (100) planes of the original f.c.c. lattice.
\bibitem{APB}We did not observe the anti-phase domains and in Ref.
\onlinecite{FullPaper} it will be shown that they are not
favored energetically.
\bibitem{MC-PhTr}Often the inflection point is used to approximate the bulk phase transition point in MC simulations finite size samples.
\bibitem{Equilibr}All of the results presented in this paper correspond to equilibrium
states and are, therefore, independent of the particular kinetic
pathways that lead to these states.
\end{thebibliography}
\end{document}